\documentstyle[prl,aps,multicol]{revtex}

\newenvironment{Fig}[1]{
\begin{figure}
\noindent\begin{minipage}[t]{\linewidth}
\begin{center}
\leavevmode
\epsfxsize=\linewidth
\epsfbox{#1}
}
{
\end{center}
\end{minipage}
\end{figure}
}

\begin{document}
\bibliographystyle{prsty}
\input epsf
\title{Generalized Compressibility in a Glass Forming Liquid}
\author{Herv\'e M. Carruzzo and Clare C. Yu}
\address{
Department of Physics and Astronomy, University of California,
Irvine, Irvine, California 92697}
\date{\today}
\maketitle
\begin{abstract}
We introduce a new quantity to probe the glass
transition.  This quantity is a linear generalized compressibility which
depends solely on the positions of the particles.  We have performed a
molecular dynamics simulation on a glass forming liquid consisting of
a two component mixture of soft spheres in three dimensions. As the
temperature is lowered, the generalized compressibility drops 
sharply at the glass transition. 
Our results are consistent with the kinetic view of the glass transition,
but not with an underlying second order phase transition.
\end{abstract}


\begin{multicols}{2}
\narrowtext The glass transition is still not well understood despite
extensive study. Experimentally the glass transition occurs when 
the relaxation time
exceeds the measurement time and particle motion appears to be arrested.
It is characterized by both kinetic and thermodynamic
features.  In the supercooled liquid kinetic quantities
such as the viscosity and relaxation time grow rapidly as the
temperature is lowered. The glass transition is also reflected in
thermodynamic quantities, e.g., the specific heat at constant pressure has a
step--like form and the dielectric constant has a peak
at a frequency dependent temperature.

Theoretically there have been two main approaches to the problem
\cite{Ediger96,Fourkas97}:
dynamic and thermodynamic. The first category has been dominated by
mode coupling theory (MCT) in which ideally the relaxation time diverges at
a temperature $T_{C}$ above the experimental glass transition \cite{Gotze92}.
The kinetic view has produced
interesting and fruitful concepts such as the influence
of the energy landscape on relaxation processes \cite{Goldstein69,Sastry98}
and dynamic inhomogeneities \cite{Donati98,Yamamoto97a,Tracht98}.
The thermodynamic viewpoint attributes the glass transition to 
an underlying phase transition hidden from direct
experimental observation by extremely long relaxation times
\cite{Ediger96,Fourkas97,Adam65,Gibbs58,Mezard99}.
In most scenarios there is an underlying second order phase
transition associated with a growing correlation length which produces
diverging relaxation times as well as diverging static susceptiblities 
\cite{Kirkpatrick89,Sethna91,Kivelson95,Ernst91,Dasgupta91,Menon95}.
More recently Mezard and Parisi \cite{Mezard99} have argued 
that the underlying transition is actually a random first order 
transition signaled by a discontinuity in the specific heat.

In an effort to better characterize the glass transition we introduce
a novel probe which we call a generalized
compressibility \cite{Yvon58}. Unlike the specific heat which monitors energy
fluctuations, this linear compressibility is a function of the
microscopic structure of the system: it depends solely on the
positions of the particles and not on their previous history. 
It is a thermodynamic quantity in the sense that it is purely
a function of the microstate of the system dictated by its location
in phase space.
It is easy to compute numerically, and it is simpler than the
dielectric constant which involves both the translation and
orientation of electric dipoles. 
By performing a molecular dynamics simulation of a
two component system of soft spheres, we find that the linear 
generalized compressibility drops sharply as the temperature
decreases below the glass transition temperature $T_g$. The drop
becomes more and more abrupt as the measurement time increases.

We now derive expressions for
the linear and nonlinear generalized compressibility. 
To probe the density fluctuations, we follow the approach
of linear response theory and consider applying an external 
potential ${\Delta P\over\rho_o}\phi(\vec r)$ which couples to the
local density $\rho (\vec r)=\sum_{i=1}^{N}\delta (\vec r - \vec r_i)$
where $\vec r_i$ denotes the position of the $i^{th}$ particle.
$\rho_o$ is the average density.
$\Delta P$ has units of pressure and sets the magnitude of the perturbation.
$\phi (\vec r)$ is a dimensionless function of
position that must be compatible with
the periodic boundary conditions
imposed on the system, i.e., it must be continuous across the
boundaries, but is otherwise arbitrary.
This adds to the Hamiltonian H of the system a term
\begin{equation}
U={\Delta P\over\rho_o}\int_Vd^3r\phi (\vec r)\rho (\vec r)
={\Delta P\over\rho_o}\sum_i\phi (\vec r_i)
\equiv {\Delta P\over\rho_o}\rho_{\phi}
\end{equation}
where we have defined 
$\rho_{\phi}=\int_Vd^3r\phi (\vec r)\rho (\vec r)=\sum_i\phi (\vec r_i)$.
$\rho_{\phi}$
is the inner product of $\phi$ and $\rho(\vec r)$, and we can regard
it as a projection of the density onto a basis function $\phi(r)$,
i.e., $\rho_{\phi}=<\rho|\phi>$. 
It weights the density fluctuations according to their spatial position. 
The application of the external potential will induce an average change 
$\delta \rho_{\phi}$ in $\rho_{\phi}$: 
\begin{equation}
\delta \rho_{\phi}=\langle\rho_{\phi}\rangle_{U}-
\langle\rho_{\phi}\rangle_{U=0}
\label{eq:deltarho}
\end{equation}
where the thermal average $\langle\rho_{\phi}\rangle_{U}$ is given by
\begin{equation}
\label{eq:AV}
\langle \rho_{\phi}\rangle_{U}={1\over {\cal Z}}\hbox{\rm Tr}
\left[ e^{-\beta (H+U)}\rho_{\phi}\right]\quad 
\end{equation} 
The partition function ${\cal Z}=\hbox{\rm Tr}e^{-\beta (H+U)}$ and
$\beta$ is the inverse temperature.
For small values of $\Delta P$, this
change can be calculated using perturbation theory.
Up to third order in $\Delta P$, we find
\begin{eqnarray}
\label{eq:expansion}
\nonumber
\delta \rho_{\phi}&=& 
        -\frac{\beta\Delta P}{\rho_o}\langle \rho_{\phi}^2\rangle_c +
        {\beta^2\Delta P^2\over 2\rho_o^2}\langle 
        \rho_{\phi}^3\rangle_c\\ 
        && -{\beta^3\Delta P^3\over 6\rho_o^3}
        \langle \rho_{\phi}^4\rangle_c,
\end{eqnarray}
where the cumulant averages are 
\begin{eqnarray}
\langle \rho_{\phi}^2\rangle_c&=&\langle \rho_{\phi}^2\rangle -
        \langle \rho_{\phi}\rangle^2 \\
\label{eq:CUMU2}
\langle \rho_{\phi}^3\rangle_c&=&\langle \rho_{\phi}^3\rangle -
        3\langle \rho_{\phi}\rangle\langle \rho_{\phi}^2\rangle +
        2\langle \rho_{\phi}\rangle^3 \\ \nonumber
\langle \rho_{\phi}^4\rangle_c&=&\langle \rho_{\phi}^4\rangle -
        4\langle \rho_{\phi}\rangle\langle \rho_{\phi}^3\rangle -
        3\langle \rho_{\phi}^2\rangle^2 + \\
        && 12\langle \rho_{\phi}\rangle^2\langle \rho_{\phi}^2\rangle -
        6\langle \rho_{\phi}\rangle^4
\end{eqnarray}
with the thermal average
$\langle \rho_{\phi}^n\rangle=\langle \rho_{\phi}^n\rangle_{U=0}$.
The third order cumulant, eq.(\ref{eq:CUMU2}), is zero in the liquid phase
because for every configuration there exists
an equivalent configuration with the opposite sign 
of $\rho_{\phi}-\langle \rho_{\phi}\rangle$ and so we will not 
consider this term any further. 
We can recast eq. (\ref{eq:expansion}) as a power
series in the perturbation $\Delta P$:
\begin{equation}
\label{eq:DV2}
{\delta\rho_{\phi} \over {N}}=
-{1\over 6\rho_o k_B T}\chi_{l}\Delta P+{1\over 6(\rho_o k_B T)^3}\chi_{nl}
(\Delta P)^3
\end{equation}
where 
\begin{equation}
\chi_{l}={6 \over N}\langle (\rho_{\phi})^2\rangle_c \quad
\chi_{nl}=-{ 1 \over N}\langle (\rho_{\phi})^4\rangle_c.
\label{eq:susceptibilities}
\end{equation}
where $k_B$ is Boltzmann's constant. In the remainder of this paper we
will focus our attention on the linear ($\chi_{l}$) and nonlinear
($\chi_{nl}$)  dimensionless generalized compressibilities defined by the
above expressions. We now discuss the choice of the function $\phi$.
We consider applying the potential along the
direction $\mu$ of one of the coordinate axes so that $\phi(\vec
r)=\phi(r^{\mu})$. A
natural candidate for $\phi(r^{\mu})$ is $\cos(k_{\mu}r^{\mu})$ with $k=2\pi
n/L$, where $n=1,2,...$ and $L^3$ is the volume $V$.
In this case, $\rho_{\phi}$ is the $k^{th}$ mode of
the cosine  transform of the density.
However, it is sufficient to consider the
simpler function $\phi(r^{\mu})=|r^{\mu}|/L$.  The absolute value
means that all the particles feel a force along the
$\mu$th direction pointing towards the origin. The results are
very similar to $\phi(r^{\mu})=\cos(k_{\mu}r^{\mu})$ for
small $k$ at a fraction of the computational cost. So our results in
this paper correspond to $\rho_{\phi}=\sum_{i}|r^{\mu}_{i}|/L$.  This
is rather like a center of mass. Since the system is isotropic, we average 
over the direction $\mu$.

We have performed a molecular dynamics simulation on a 
glass forming liquid \cite{Weber85,Kob94} consisting of a 50:50
binary mixture of soft spheres in three dimensions.
The two types of spheres, labelled A and B, differ only in their sizes.
The interaction between two particles a distance
$r$ apart is given by $V_{\alpha\beta}(r)
=\epsilon[(\sigma_{\alpha\beta }/r)^{12}+X_{\alpha\beta}(r)]$
where the interaction length
$\sigma_{\alpha\beta }=(\sigma_{\alpha}+\sigma_{\beta})/2$,
$\sigma_B/\sigma_A=1.4$ ($\alpha$, $\beta =$ A, B). 
For numerical efficiency, we set
the cutoff function $X_{\alpha\beta}(r)=r/\sigma_{\alpha\beta}-\lambda$
with $\lambda =13/12^{12/13}$. The interaction is cutoff at
the minimum of the potential $V_{\alpha\beta}(r)$. 
Energy and length are measured in units of 
$\epsilon$ and $\sigma_A$, respectively. Temperature is given in units of
$\epsilon /k_B$, 
and time is in units of $\sigma_A\sqrt{m/\epsilon}$ where
$m$, the mass of the particles, is set to unity. 
The equations of motion were integrated using the leapfrog
method \cite{Rapaport95} with a time step of 0.005.
During each run the temperature was kept constant using a constraint
algorithm \cite{Rapaport95}. $N=N_A+N_B$ is the total number of particles. 
The system occupies a cube with dimensions 
($\pm$ L/2, $\pm$ L/2, $\pm$ L/2) and periodic boundary conditions.
Since $N$ and $L$ are fixed in any given run, the density
$\rho_o=N/L^3$ is also fixed.
\begin{Fig}{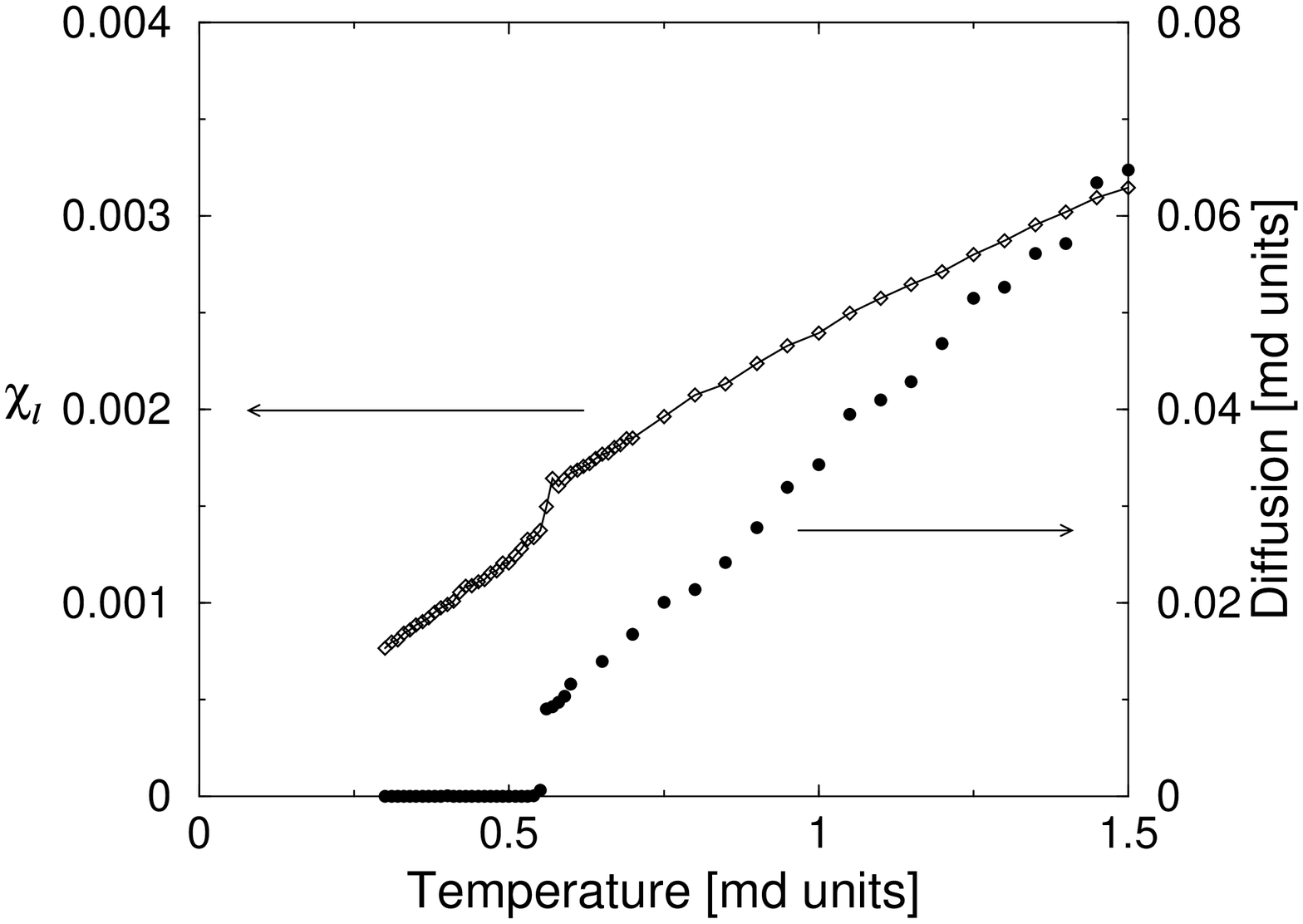}
\caption{Linear generalized compressibility and diffusion constant of a one
component system as a function of temperature for 512 soft
spheres. Crystallization is clearly seen around T=0.57 ($\rho_o
=1.1$). The measurement time was $10^6$ md steps for each
temperature. The compressibility was averaged over 5 runs while the
diffusion is shown for a single run.}
\label{fig:crystal}
\end{Fig}

As a point of reference we determined the mode coupling
$T_C$ by fitting the data for the relaxation time
$\tau(T)$ which is defined as the
time when the self part of the intermediate scattering function
$F_{s,\alpha}(\vec{k},t)$ falls to $1/e$ \cite{Kob94}.
\begin{equation}
F_{s,\alpha}(\vec{k},t)=\frac{1}{N_{\alpha}}\left\langle
\sum_{i=1}^{N_{\alpha}}e^{i\vec{k}\cdot(\vec{r}_{i}(t)-\vec{r}_{i}(0))}
\right\rangle
\label{eq:SelfFkt}
\end{equation}
where the subscript $\alpha$ refers to the particle type, A or B.
We choose B particles.
$\vec{r}_{i}(t)$ is the position of particle $i$ at time $t$,
and $\langle ...\rangle$ refers to an average over different configurations.
The wavevector $\vec{k}=2\pi\vec{q}/L$ where $\vec{q}$ is a vector
of integers. For an isotropic system $F_{s,\alpha}(\vec{k},t)$ depends
only on the magnitude $k=|\vec{k}|$. We choose $k=k_{max}=2\pi q_{max}/L$
where $k_{max}$ is the position of the first maximum of the partial static
structure factor $S(k)$. For B type particles we use $q_{max}=8.3666$.
We fit the ideal
MCT form \cite{Gotze92} $\tau(T)= A^{\prime}(T-T_{C})^{-\gamma}$
with $A^{\prime}=69$, $\gamma=1.6$, $T_{C}=0.306\pm 0.005$.
In finding $T_C$, we used data from 7 temperatures
in the range $0.33<T\leq 0.39$ below the caging temperature
($T_{\rm cage}=0.4$) where the intermediate scattering function
first begins to show a plateau. Our fit is consistent with the
mode coupling theory requirement that $\gamma\geq 1.6$.

Our procedure for doing runs is as follows. We start each run at a high
temperature (T=1.5) and lower the temperature in steps of $\Delta
T=0.05$. At each temperature we equilibrate for $10^4$ molecular
dynamics (md) steps and then measure the quantities of interest
for $N_{\tau}$ additional steps where $N_{\tau}=10^5$, $2\times
10^5$, $10^6$, $3\times 10^6$ or $10^7$.  All the particles move at each md
step. The results are then averaged over up to 40 different initial
conditions (different initial positions and velocities of the spheres).  

As a check on our procedure for measuring $\chi_{l}$ and
$\chi_{nl}$, we consider first the case of the crystal. To this
end, we consider a system of 512 identical  ($\sigma_A=\sigma_B$)
particles at a density $\rho_o=1.1$.
Figure \ref{fig:crystal} shows the linear compressibility and the
diffusion constant as a function of temperature for the single
component liquid.  At high temperatures $\chi_{l}$ has a small
slope which becomes steeper at low temperature. The salient
feature is the very sharp drop around T=0.57 of the linear generalized
compressibility and the diffusion constant. The specific heat, which is
not shown, has a sharp delta function--like peak around T=0.57. The low
temperature phase (T$<$0.57) is a crystal with sharp Bragg peaks in
the structure factor. Upon
heating and cooling, the transition shows hysteresis. All these
observations are consistent with the fact that crystallization is a
first order transition.  Not shown here is the nonlinear
compressibility which is zero within our numerical error. 

We now examine the response of the two component glass forming
liquid. All the runs were done at a density of $\rho_o=0.6$ and
$\sigma_B/\sigma_A=1.4$. For these parameters  crystallization is
avoided upon cooling.
\begin{Fig}{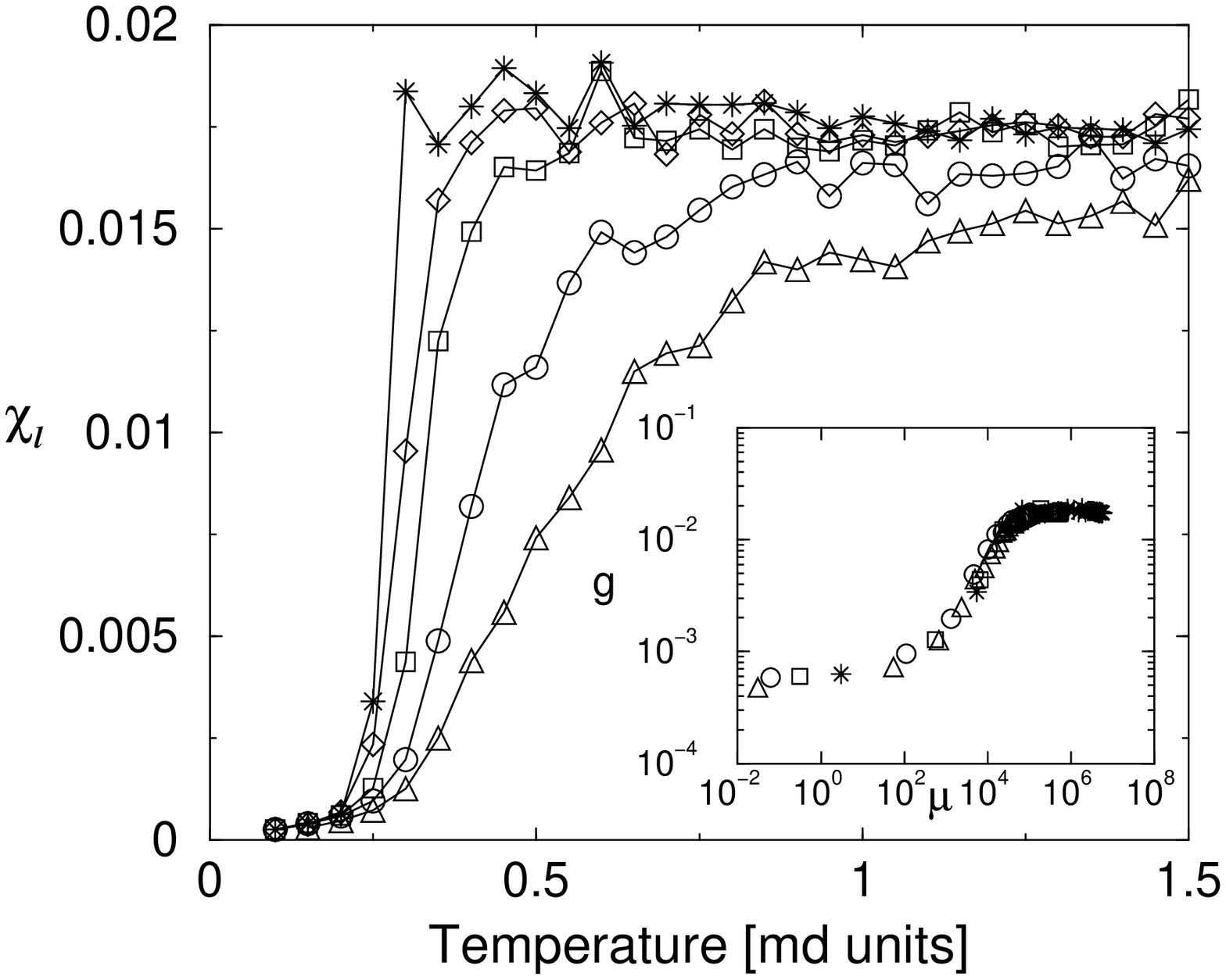}
\caption{Linear generalized compressibility as a function of
temperature for different measuring times $t_M$: $10^5$ ($\triangle$, 40
runs), $2\times 10^5$ ($\circ$, 32 runs), $10^6$ ($\Box$, 10 runs), 
$3\times 10^6$ ($\Diamond$, 6 runs) and $10^7$ ($\ast$, 4 runs)
md steps. System size is 512
particles. $\rho_o = 0.6$ and  $\sigma_B/\sigma_A=1.4$. Inset: 
$T>T_o$ subset of the same data scaled as described in text.}
\label{fig:glass_lin_susc}
\end{Fig}
Figure \ref{fig:glass_lin_susc} shows the linear generalized
compressibility as a function of temperature for different run times.
The compressibility at high temperatures is independent of $T$ and
about an order of magnitude larger than that of the single component
fluid. In the vicinity of the glass transition $\chi_{l}$
drops. Notice that as the measuring time $t_M$ increases, the temperature
of the drop decreases and becomes more abrupt.
The linear compressibility is proportional to the width of the
distribution of $\rho_{\phi}$. If we regard $\rho_{\phi}$ as a
generalized center of mass, then the drop in $\chi_{l}$ corresponds to
the sudden narrowing of the distribution $P(\rho_{\phi})$ and the
sudden arrest in the fluctuations of $\rho_{\phi}$.  This behavior can
be quantified using a scaling ansatz:
$\chi_{l}(t_M,T)=g(\mu=t_M/\tau (T))$, where the characteristic time
has the Vogel--Fulcher form $\tau(T)=\exp (A/(T-T_o))$. The inset
of Figure \ref{fig:glass_lin_susc} shows that the data collapse onto
a single curve with $A=0.75$, $T_o=0.15$. (The data could not be
fitted using $\tau(T)=\tau_{MCT}(T)=A^{\prime}(T-T_C)^{\gamma}$ as suggested 
by simple mode
coupling theories \cite{Gotze92}.) This value of $T_o$ lies below
the mode coupling $T_{C}=0.306$, the upper bound
of the glass transition temperature. 
Notice in Fig. \ref{fig:glass_lin_susc} that the drop associated
with 10$^7$ md steps occurs approximately at the mode coupling 
$T_C$, indicating that our long runs were in equilibrium down to the
mode coupling transition temperature.  This scaling suggests that
$\chi_{l}$ becomes a step function for infinite $t_M$ and that
the drop in the
compressibility would become a discontinuity at infinitely long times.
This is consistent with a sudden arrest of the motion of the
particles in the liquid which is the kinetic view of the glass
transition. The abrupt drop also appears to be
in agreement with Mezard and Parisi's proposal that the glass
transition is a first order phase transition \cite{Mezard99}. 
However, the drop indicates that our simulations are falling out
of equilibrium and therefore we cannot really tell if there is a true
thermodynamic transition.

As an independent check of the glass transition temperature, we have 
calculated the specific heat which is shown in Figure \ref{fig:spht}.
Notice that the temperature of the peak in the specific heat agrees 
with the temperature at which $\chi_{l}$ drops.
\begin{Fig}{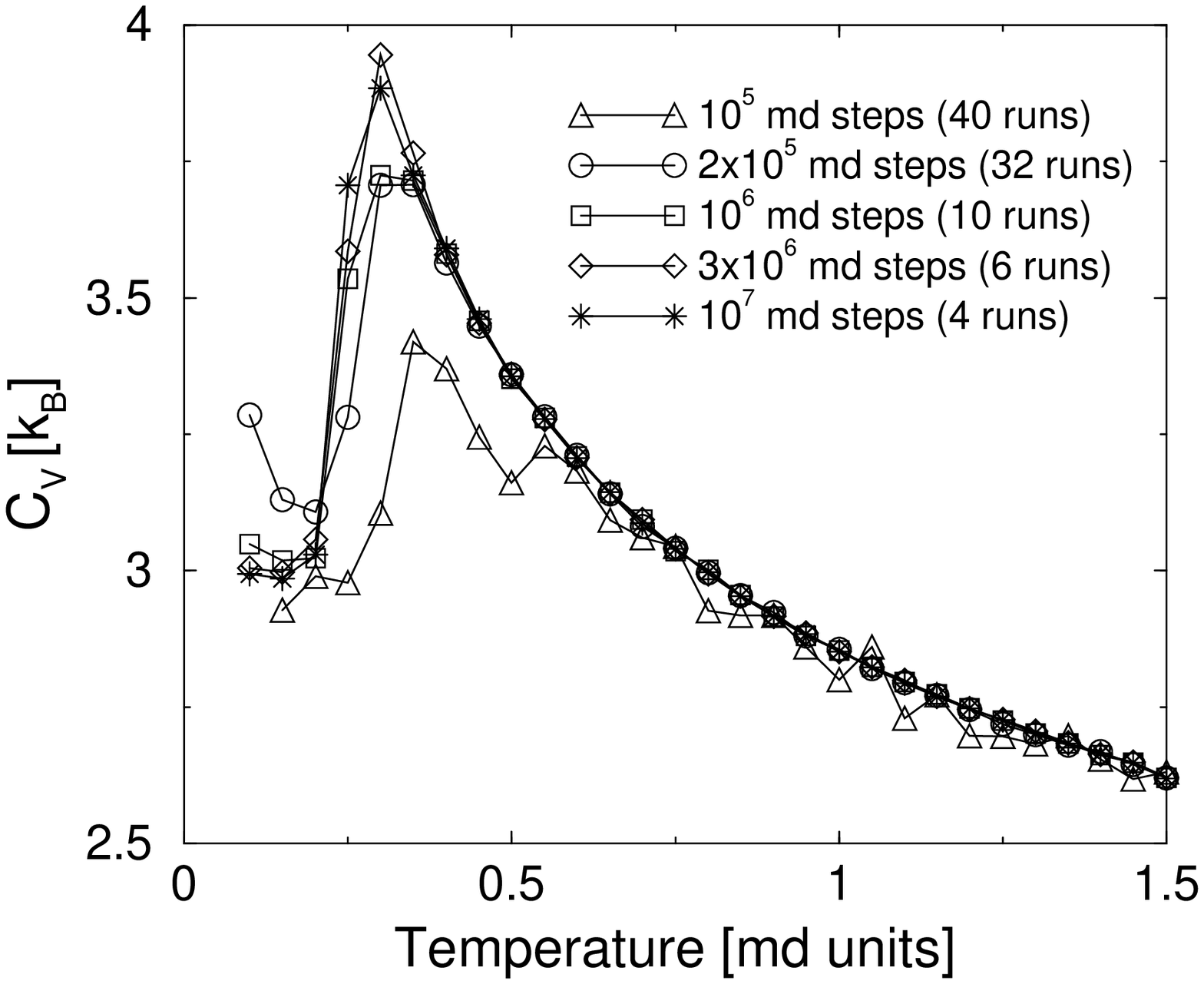}
\caption{Specific heat $C_V$ at constant volume as a function of temperature
calculated using energy fluctuations.
All parameters are the same as in Figure \protect\ref{fig:glass_lin_susc}.}
\label{fig:spht}
\end{Fig}

The behavior seen in Figure \ref{fig:glass_lin_susc} is similar to
that seen in measurements of the real part of the frequency dependent
dielectric function $\varepsilon^{\prime}(\omega)$ \cite{Menon95}. In
that case as the frequency decreased, the temperature of the peak in
$\varepsilon^{\prime}(\omega)$ decreased and the drop in
$\varepsilon^{\prime}(\omega)$ below the peak became more abrupt.  By
extrapolating their data to $\omega=0$,  Menon and Nagel \cite{Menon95}
argued that $\varepsilon^{\prime}(\omega=0)$ should diverge at the glass
transition, signaling a second order phase  transition. We have looked
for evidence of this divergence by examining samples of different
sizes to see if the linear generalized
compressibility increased systematically
with system size. As shown in Figure \ref{fig:glass_size_dep} we find
no size dependence and no indication of a diverging linear generalized
compressibility. We also find no size dependence for the specific
heat and hence, no evidence of a diverging specific heat (not shown). 
This is corroborated by recent MCT calculations of a molten salt
which find that $\varepsilon^{\prime}(\omega\rightarrow 0)$ goes to a
finite value as the glass transition is approached \cite{Wilke99}.

We have found hysteresis at the glass transition
by first cooling a system of 512 particles to our lowest temperature
$T=0.1$ and then heating in steps of $\Delta T=0.05$. 
As before we equilibrate at each temperature for
$10^4$ time steps and then measure quantities for an additional $10^6$
time steps. Our results are shown in the inset of
Figure \ref{fig:glass_size_dep}.
Notice the slight hysteresis with the rise in $\chi_{l}$ upon
warming being at a slightly higher temperature than the drop in
$\chi_{l}$ upon cooling. This hysteresis is consistent with 
the kinetic arrest of motion. 
\begin{Fig}{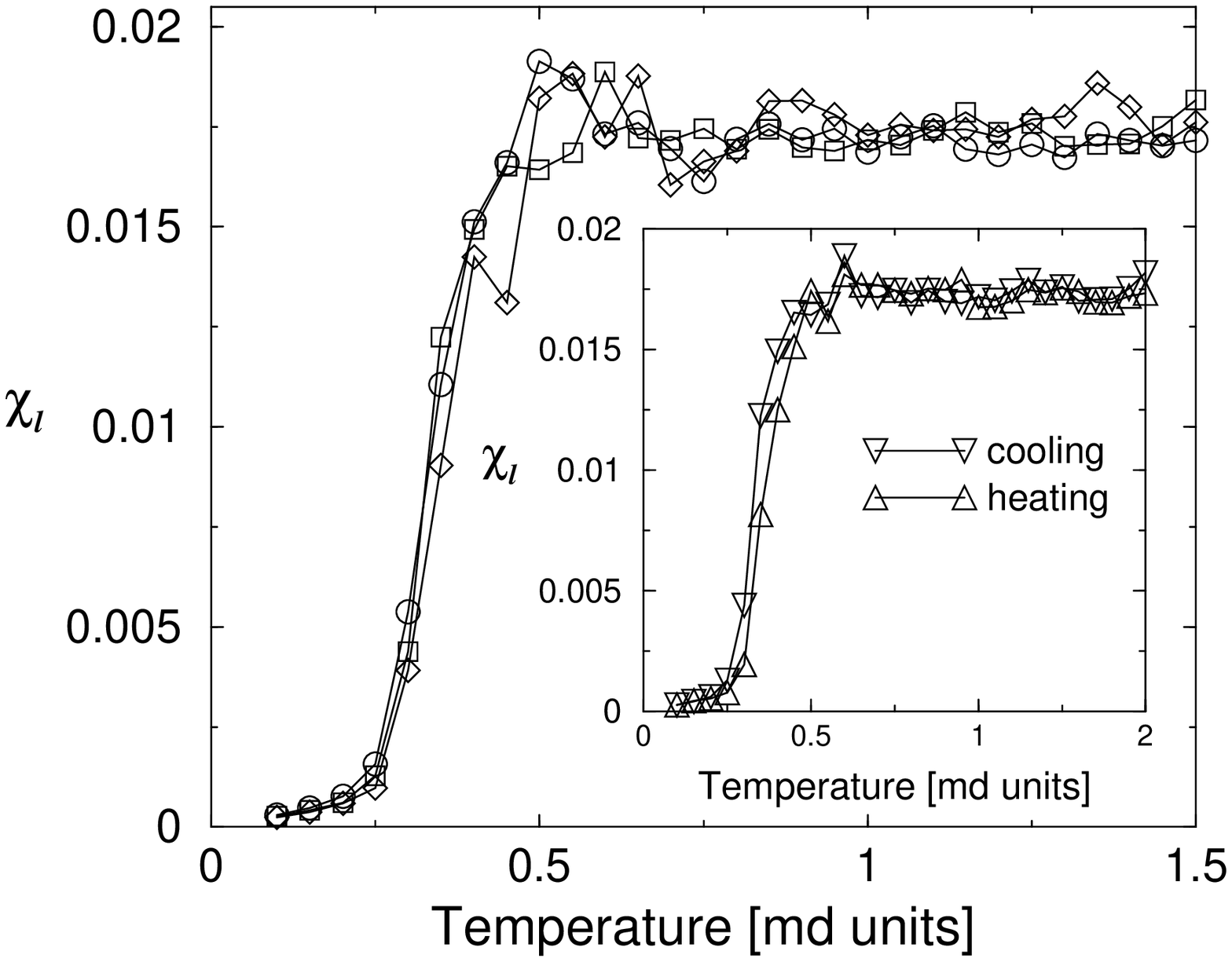}
\caption{Linear compressibility as a function of temperature for
different system sizes: 216 ($\circ$, 5 runs), 512 ($\Box$, 10 runs),
and 1000 ($\Diamond$, 5 runs) particles. The measuring time
was $10^6$ md steps in all cases. Inset: Linear generalized
compressibility as a function of temperature for system
of 512 particles upon cooling and heating. The measuring time was
$10^6$ md steps in both cases. The data was averaged over 10 runs.
Other parameters for the main figure and the inset
are the same as in Figure \protect\ref{fig:glass_lin_susc}.}
\label{fig:glass_size_dep}
\end{Fig}

We now turn to the case of the nonlinear generalized compressibility
$\chi_{nl}$ given by eq. (\ref{eq:susceptibilities}). We are
motivated by the case of spin glasses where 
the nonlinear magnetic compressibility diverges at the spin glass
transition while the linear compressibility only has a cusp 
\cite{Bhatt88,Levy86}.
There have been a few studies of nonlinear response functions in real
glasses \cite{Dasgupta91,Wu91}, but these have not found any
divergences.  Our results are consistent with this conclusion. In
particular we find that the nonlinear generalized compressibility is
zero above and below the glass transition temperature, though it does show a
glitch at the glass transition. There is no systematic increase
with system size, indicating the absence of a divergence. 
Because $\chi_{nl}$ is sensitive to
the tails of the distribution of $\rho_{\phi}$, one must be careful to
obtain a good ensemble average in the liquid above the glass
transition temperature.  We have done this by doing 32 runs, each
involving 200,000 time steps, with different initial conditions,
stringing them together as though they were one long run and then
taking the appropriate averages. This produces a better ensemble
average of $<\rho_{\phi}^2>^2$ which enters into $\chi_{nl}$ in
eq. (\ref{eq:CUMU2}). $\chi_{nl}$ also took longer to equilibrate
than $\chi_{l}$. A plot $\chi_{nl}$ versus run time shows that 
one needs to run at least $10^6$ time steps before
$\chi_{nl}$ appeared to saturate.

To summarize, we have introduced a new thermodynamic quantity which
depends solely on the positions of the particles and not on their
histories. This quantity drops abruptly at the glass transition which  
is compatible with a
kinetic arrest of motion, but not with an underlying second order
phase transition. This generalized compressibility can be 
measured experimentally. It can be directly measured in colloidal
experiments which monitor the positions of the particles \cite{Weeks00}. 
Measurements of the width of the distribution of
$\rho_{q}$, the spatial Fourier transform of the density, would also
give the linear generalized compressibility.
 
We thank Andrea Liu and Sharon Glotzer for helpful discussions.
This work was supported in part by 
CULAR funds provided by the University of California for the conduct of
discretionary research by Los Alamos National Laboratory.

\end{multicols}
\end{document}